\documentclass[amsmath,amssymb, aps, prl, superscriptaddress,twocolumn]{revtex4-1}
\usepackage[pdftex,colorlinks=true]{hyperref}
\hypersetup{
citecolor = blue
}

\usepackage{graphicx}
\usepackage{dcolumn}
\usepackage{bm}
\usepackage{amsmath}
\usepackage{float}
\usepackage{hyperref}
\usepackage{wrapfig}
\usepackage{amssymb}
\usepackage{enumitem}
\usepackage{subfigure}
\usepackage{placeins}

\usepackage{amsfonts}
\usepackage{upgreek}
\usepackage{slashed}
\usepackage{latexsym}

\usepackage[normalem]{ulem}
\usepackage{amsthm}
\usepackage{bbm}
\usepackage{array}

\newcommand{\beq}{\begin{equation}}
\newcommand{\eeq}{\end{equation}}
\newcommand{\bmeq}{\begin{multline}}
\newcommand{\emeq}{\end{multline}}
\newcommand{\nn}{\nonumber \\}

\def \dg{\dagger}

\begin{document}

\title{Coherent superconductivity with a large gap ratio from incoherent metals}

\author{Aavishkar A. Patel}
\affiliation{Department of Physics, Harvard University, Cambridge MA 02138, USA}
\affiliation{Kavli Institute for Theoretical Physics, University of California, Santa Barbara CA 93106-4030, USA}

\author{Michael J. Lawler}
\affiliation{Department of Physics, Cornell University, Ithaca, NY 14853, USA}
\affiliation{Department of physics, Binghamton University, Vestal, NY 13850, USA}

\author{Eun-Ah Kim}
\affiliation{Department of Physics, Cornell University, Ithaca, NY 14853, USA}

\date{\today}

\begin{abstract}
A mysterious incoherent metallic (IM) normal state with $T$-linear resistivity is ubiquitous among strongly correlated superconductors. Recent progress with microscopic models exhibiting IM transport has presented the opportunity for us to study new models that exhibit direct transitions into a superconducting state out of IM states within the framework of connected Sachdev-Ye-Kitaev (SYK) ``quantum dots".  Here local SYK interactions within a dot produce IM transport in the normal state, while local attractive interactions drive superconductivity. Through explicit calculations, we find two features of superconductivity arising from an IM normal state: First, despite the absence of quasiparticles in the normal state, the superconducting state still exhibits coherent superfluid transport. Second, the non-quasiparticle nature of the IM Green's functions produces a large enhancement in the ratio of the zero-temperature superconducting gap $\Delta$ and transition temperature $T_{sc}$, $2\Delta/T_{sc}$, with respect to its BCS value of $3.53$. 

\end{abstract}
\maketitle

Superconductivity in correlated systems often emerges from a mysterious incoherent metallic (IM) state with $T$-linear resistivity. The origin of the $T$-linear resistivity has been a subject of active research and debate~\cite{Varma-PhysRevLett.63.1996,Hartnoll-NP14,Hartnoll-Karch-PhysRevB.91.155126,Bruin804}. Moreover Refs.~\cite{emery-kivelson1995-nature,Emery-Kivelson95PRL-PhysRevLett.74.3253,Zaanen-Plankian} have pointed out that superconductivity emerging out of such strange metals should be qualitatively different from that emerging out of conventional metals. 
Nevertheless the lack of a solvable microscopic model has prevented the community from forming a concrete connection between many inexplicable properties of the superconducting state and the IM state in correlated systems. 

Recent proposals of microscopic models exhibiting IM transport in a solvable limit~\cite{Gu2017,Sachdev2017,Balents2017,Zhang2017,Patel2017,Chowdhury2018,Wu2018}, present new avenues. The approach shared among these models is to build on \textcite{Sachdev-Ye} finding non-Fermi liquid Green's functions in a solvable model of fermions with infinite-range interactions. Although both this original model and a simpler model with Majorana fermions~\cite{Kitaev} exhibit non-Fermi liquid properties as well as interesting connections to quantum gravity~\cite{Kitaev,Sachdev2015} in the solvable limit, they do not support local current operators. However, by introducing local coupling between multiple copies of these infinite ranged models, in the spirit of weakly coupled quantum dots each hosting multiple orbitals, Refs.~\cite{Zhang2017,Gu2017,Sachdev2017,Balents2017,Patel2017,Chowdhury2018} established solvable microscopic models with IM transport. These models led to new insights regarding loss of quasiparticle coherence during scattering leading to such transport. But moreover, they have put us in an opportune moment to theoretically study the properties of superconducting (SC) phases born out of such IMs, in a solvable limit.

In this work, we consider two models that can be solved in a large-$N$ limit that demonstrate the much sought after transition from an IM with $T$-linear resistivity to SC. We then study the implication of strong correlations destroying coherent quasiparticles on the superconducting transition and state. In spite of the incoherent normal state, the paired state still supports a coherent supercurrent. We further show that a key prediction of the Bardeen-Cooper-Schrieffer (BCS) mean-field theory~\cite{BCS} of superconductivity is violated: the ratio between the zero temperature gap $\Delta$ and the transition temperature $T_{sc}$ far exceeds the BCS value of $2\Delta/T_{sc}\approx 3.53$. We compare this mechanism of gap ratio enhancement with that in the Eliahsberg theory and in experiments.

\begin{figure}
\includegraphics[width=.48\textwidth]{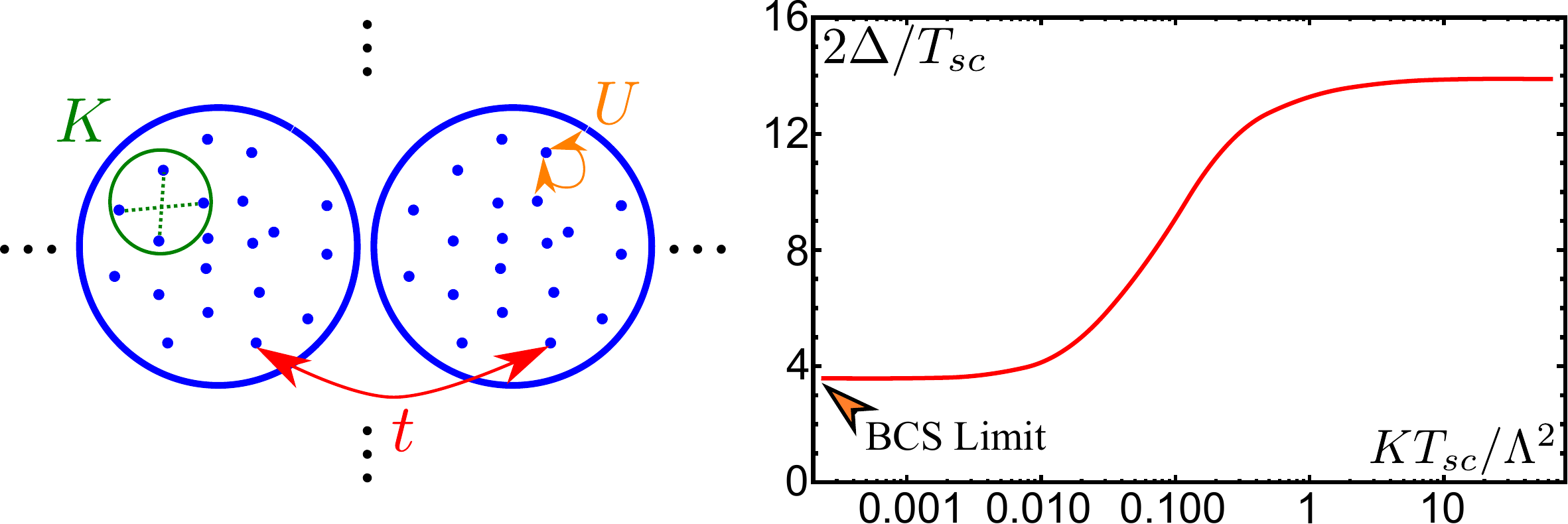}
\caption{{\bf Left:} A cartoon of Model 1. {\bf Right:} A plot of $2\Delta/T_{sc}$ as $T\rightarrow0$, vs $KT_{sc}/\Lambda^2$ for $K=1000$, $T_{sc}=10$, for different values of the bandwidth $\Lambda\sim t$ in Model 1. The value of $U$ is adjusted as $\Lambda$ is varied in order to keep $T_{sc}$ fixed. For large $\Lambda$, the transition to SC is from a dispersive Fermi liquid, and we find the BCS result $2\Delta\approx 3.53T_{sc}$. For small values of $\Lambda$, such that $KT_{sc}/\Lambda^2\gg1$, the transition to SC is from a non-Fermi liquid IM with a linear-in-$T$ resistivity, and $2\Delta\gg 3.53T_{sc}$.}
\label{fig:GapTc42}
\end{figure}

{\it Model 1} --
We consider a lattice model of two species of fermions $a,b$ with disordered local on-site Sachdev-Ye-Kitaev  interactions of 4th order (SYK$_4$), but with a {\it uniform} quadratic hopping, and an attractive term that pairs the two species locally (Fig.~\ref{fig:GapTc42}). It is given by
\begin{align}
&H_1 = \sum_m \sum_{i_1,..,i_4=1}^{N}\left(K^{a m}_{i_1,..,i_4}a^\dg_{i_1m}a^\dg_{i_2m}a_{i_3m}a_{i_4m} + (a\leftrightarrow b) 
\right) \nn
&-t\sum_{\langle mn\rangle} \sum_{i=1}^{N}\left(a^\dg_{im}a_{in} + 
(a\leftrightarrow b)
+\mathrm{H.c.}\right) \nn
&-\frac{U}{N}\sum_m \sum_{i,j=1}^N b^\dg_{im}a^\dg_{im}a_{jm} b_{jm},
\label{H42}
\end{align}
where $m$ and $n$ are the site indices with $N$ fermions of each type.
Here the disordered complex Gaussian random couplings $K^{\alpha m}_{i_1,..,i_4}$ satisfy  $\ll K_{i_1,i_2,i_3,i_4}^{\alpha m}K_{i_4,i_3,i_2,i_1}^{\beta n}\gg=K^2/(8N^3)\delta_{\alpha\beta}\delta_{mn}$,
where $\ll~..~\gg$ denotes disorder-averaging, and all other averages are zero. A simpler model, without the attractive $U$ term, and with only one type of fermion, was first proposed in Ref.~\cite{Zhang2017}.

Without the $U$ term, and with $K\gg t$, this model exhibits a crossover between a high temperature IM state with $T$-linear resistivity to a low tempterature Fermi liquid state. Specifically, for $K\gg T\gg t^2/K$, the 
Green function is asymptotically given by the local SYK Green's function in imaginary time~\cite{Sachdev2015}:
\beq
G^{a,b}(0<\tau< \beta) = - \frac{\pi^{1/4}}{K^{1/2}\sqrt{2}}\left(\frac{T}{\sin(\pi T \tau)}\right)^{1/2}.
\label{GSYK4}
\eeq
Using the Kubo formula, one can derive the linear-in-$T$ resistivity in the large-$N$ limit from the scaling of the above Green's function~\cite{Balents2017,Patel2017,Chowdhury2018}. On the other hand, for $T\ll t^2/K$, the Green function approaches that of a Fermi liquid whose resistivity scales as $T^2$~\cite{Zhang2017,Chowdhury2018}.

The attractive $U$ term leads to a spatially uniform $s$-wave pairing instability at $T=T_{sc}$. Once SC is established, the order parameter $\Delta_0 = \langle\sum_i a_{im}b_{im}\rangle/N$ condenses. In the large-$N$ limit, we then get the Dyson and gap equations (Supplementary Information)
\begin{align}
&\mathcal{G}(i\omega_n) = \int\frac{d^dk}{(2\pi)^d}\frac{G(i\omega_n,k)}{1+U^2|\Delta_0|^2|G(i\omega_n,k)|^2},\nn
&\Sigma(\tau-\tau^\prime) = -K^2\mathcal{G}^2(\tau-\tau^\prime)\mathcal{G}(\tau^\prime-\tau), \nn
&G^{-1}(i\omega_n) = i\omega_n -\xi_k - \Sigma(i\omega_n), \nn
&T\sum_{\omega_n}\int \frac{d^dk}{(2\pi)^d} \frac{|G(i\omega_n,k)|^2\Delta_0}{1+U^2 |\Delta_0|^2 |G(i\omega_n,k)|^2} = \frac{\Delta_0}{U},
\label{DysonGap42}
\end{align}
which can be iterated numerically starting with an infinitesimal $\Delta_0$ and the free fermion $G(i\omega_n)=(i\omega_n)^{-1}$ in order to determine both $G$ and $\Delta_0$.
Here $\mathcal{G}(\tau)=\mathcal{T}_\tau\langle \sum_i a_{im}(\tau)a^\dg_{im}(0)\rangle/N=\mathcal{T}_\tau\langle \sum_i b_{im}(\tau)b^\dg_{im}(0)\rangle/N$ is the local time-ordered Green's function, and $\xi_k$ is the dispersion of the fermions.

For simplicity, while still capturing the essential physics, we consider $d=2$, and $\xi_k = \Lambda k^2/(4\pi)-\Lambda/2\equiv\epsilon_k - \Lambda/2$, with $\epsilon_k\in[0,\Lambda]$, where $\Lambda\sim t$ is the bandwidth of the dispersion. We can then replace $\int\frac{d^2k}{(2\pi)^2}\rightarrow\frac{1}{\Lambda}\int_{-\Lambda/2}^{\Lambda/2}d\xi_k$, and perform all momentum integrals analytically. In general, we determine $T_{sc}$ numerically by taking the limit $\Delta_0\rightarrow0$ in the last line of (\ref{DysonGap42}). For $T>T_{sc}$, the solution of (\ref{DysonGap42}) with $\Delta_0=0$ corresponds to a stable local minimum of the free energy. The large-$N$ limit strongly suppresses fluctuations of $\Delta_0$ out of this minimum. As $T$ is lowered below $T_{sc}$, the curvature of the free energy as a function of $\Delta_0$ at $\Delta_0=0$ changes sign, and the system condenses to a new minimum with $\Delta_0\neq0$ (Supplementary Information).

When $U$ is infinitesimal so that $T_{sc}\ll t^2/K$, the SC arises out of a Fermi liquid, and we find the standard BCS result of $2\Delta=3.53T_{sc}$. On the other hand if both $K$ and $U$ are large such that $K\gg T_{sc} \gg t^2/K$, we obtain the transition to SC from the linear-in-$T$ IM. From (\ref{DysonGap42}) and (\ref{GSYK4}) we get (Supplementary Information)
\beq
T_{sc} \approx \frac{2K}{\pi}\tan^{-1}\left(e^{-\pi^{1/2}K/U}\right),
\label{Tc42}
\eeq
where we employed a UV frequency cutoff $\sim K$ in (\ref{DysonGap42}). By solving (\ref{DysonGap42}) numerically, we can study how the gap ratio evolves through the crossover between SC emerging from a Fermi liquid to  SC emerging from a linear-in-$T$ IM. For this, we study the variation of the zero-temperature gap to single-particle excitations, $\Delta$, with the bandwidth $\Lambda$, keeping $K$ and $T_{sc}$ fixed. $\Delta$ corresponds to the location of the peak of the spectral function $\mathcal{A}(\omega,\{k:\xi_k=0\})$, with
\beq
\mathcal{A}(\omega,k)=-2\mathrm{Im}\left[\frac{G_R(\omega,k)}{1+U^2|\Delta_0|^2G_R(\omega,k)G_R^\ast(-\omega,k)}\right],
\eeq 
which may be obtained from a numerical solution of the real-time version of (\ref{DysonGap42}) (Supplementary Information). 
For small interactions and $T_{sc}$ (relative to the bandwidth), SC emerges from a Fermi liquid, and the gap ratio is consistent with the ``BCS'' value of $2\Delta\approx3.53T_{sc}$. However,  the interactions $U$ and $K$ are cranked up relative to the bandwidth so that  SC emerges directly from a $T$-linear IM, the gap ratio substantially exceeds the BCS gap ratio (Fig.~\ref{fig:GapTc42}). When the gap ratio is enhanced significantly, we also find that the superconducting transition becomes first order, {\it i.e.}, the order parameter $\Delta_0$ jumps discontinuously to a nonzero value at $T=T_{sc}$. First-order transitions have also been noted earlier in studies of superconductivity arising from non-Fermi liquids~\cite{Chubukov2003,Giannakis2004}. 

{\it Model 2} -- We now consider a model that realizes an instability to SC from a non-Fermi liquid even for infinitesimal values of $U$.  
In order to avoid a Fermi liquid state as $T\rightarrow0$, we need a model that has no quadratic terms in its Hamiltonian. We however still want a linear-in-$T$ resistivity above some small temperature scale. We thus replace the on-site interactions of Model 1 with higher order SYK$_8$ terms, and the quadratic hopping between adjacent sites by pair hoppings that realize SYK$_4$ interactions between adjacent sites. As we shall explain in detail, the scaling dimensions of the current operator and the local Green's functions then lead to a linear-in-$T$ resistivity above a certain temperature. Since the charge transfer between sites is now strongly disordered, we have to use an attractive interaction given by not a conventional on-site pairing term, but rather a spatially uniform term that simultaneously binds $a-b$ pairs on site and hops them between nearest-neighbor sites, which allows coherent pair hopping below $T_{sc}$ and hence establishes superfluid phase coherence in the SC state. 

We start with a single-site $\mathrm{SYK}_8$ model with two species of fermions:
\begin{multline}
H_{2,0} = \sum_{i_1,..,i_8=1}^{N}\bigg(J^a_{i_1,..,i_8}a^\dg_{i_1}..a^\dg_{i_4}a_{i_5}..a_{i_8} +\\ J^b_{i_1,..,i_8}b^\dg_{i_1}..b^\dg_{i_4}b_{i_5}..b_{i_8}\bigg),
\label{singleH}
\end{multline}
with complex Gaussian random couplings $J^{\alpha}_{i_1,..,i_8}$ satisfying  $\ll J_{i_1,..i_8}^\alpha J_{i_1,..i_8}^{\alpha\ast}\gg=J^2/(2304N^7)$, with all other averages being zero. For $J\gg T$, the resulting $\mathrm{SYK}_8$ Green's function is~\cite{Sachdev2017} 
\begin{align}
&G^{a,b}(0<\tau<\beta) = - \frac{C_8}{J^{1/4}}\left(\frac{T}{\sin(\pi T \tau)}\right)^{1/4}, \nn
&~~C_8 = \sin^{1/4} \left(\frac{\pi }{8}\right) \Gamma^{1/8} \left(\frac{1}{4}\right) \Gamma^{1/8} \left(\frac{7}{4}\right).
\label{GSYK8}
\end{align}

\begin{figure}
\includegraphics[width=.48\textwidth]{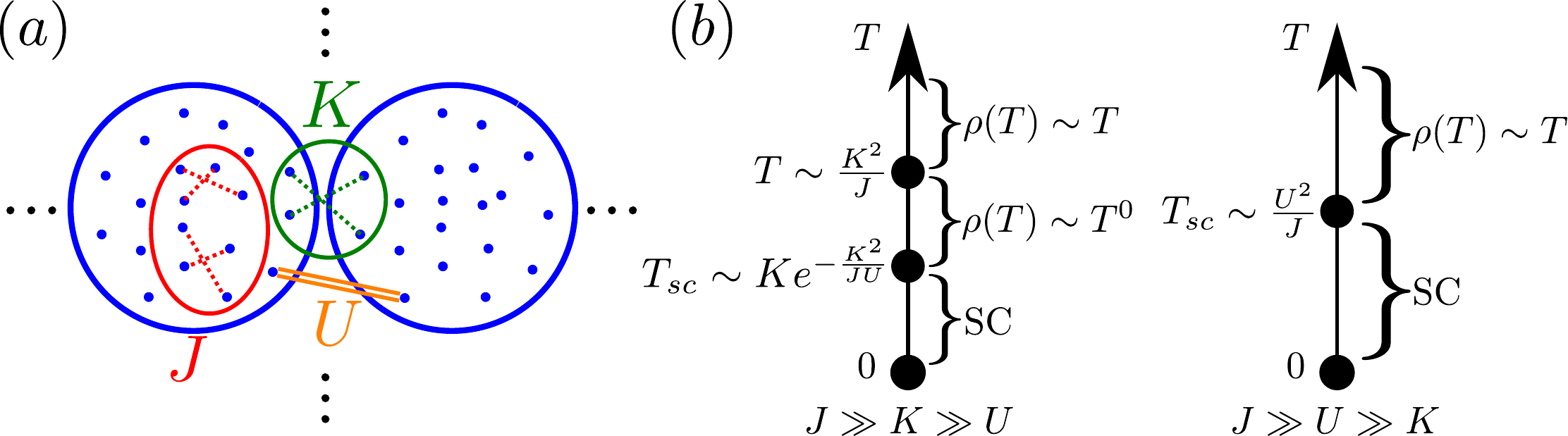}
\caption{{\bf (a)} A schematic representation of Model 2. {\bf (b)} Crossover diagrams in different regimes.  For $J\gg K\gg U$, there is first a crossover to an IM with $T$-independent resistivity, before the SC transition. For $J\gg U\gg K$, we just have a transition from an IM with $T$-linear resistivity to an SC. }
\label{fig:model-PD}
\end{figure}

We then place a system described by (\ref{singleH}) on each site of a lattice indexed by $m$. The random $\mathrm{SYK}_8$ couplings are not correlated between sites. We introduce two inter-site terms between nearest-neighbor sites: a random $\mathrm{SYK}_4$ interaction that hops $a$ and $b$ fermions independently in pairs between nearest-neighbor sites  which is necessary for IM transport, and a uniform hopping term for $a-b$ Cooper pairs  which drives superfluid phase coherence below $T_{sc}$ (Fig.~\ref{fig:model-PD}(a)):
\begin{align}
&H_2 = \sum_m \sum_{i_1,..,i_8=1}^{N}\left(J^{am}_{i_1,..,i_8}a^\dg_{i_1m}..a^\dg_{i_4m}a_{i_5m}..a_{i_8m} +(a\leftrightarrow b) 
\right) \nn
&+\sum_{\langle mn\rangle} \sum_{i_1,..,i_4=1}^{N}\Big(K^{a\langle mn\rangle}_{i_1,..,i_4}a^\dg_{i_1m}a^\dg_{i_2m}a_{i_3n}a_{i_4n} + 
(a\leftrightarrow b) \nn
&+\mathrm{H.c.}\Big)-\frac{U}{zN}\sum_{\langle m,n\rangle}\sum_{i,j=1}^N\left(b^\dg_{im}a^\dg_{im}a_{jn} b_{jn}+\mathrm{H.c.}\right),
\label{fullH}
\end{align}
with $\ll K^{\alpha \langle m,n\rangle}_{i1,i2,i3,i4},K^{\alpha \langle m,n\rangle\ast}_{i1,i2,i3,i4}\gg = K^2/(8zN^3)$, and all other averages are zero. Note that the role of the SYK$_4$ inter-site interactions in this model is distinct from the intra-site SYK$_4$ interactions in Model 1. The coordination number of the regular lattice is $z$. 

{\it Normal state of Model 2} --
For $T>T_{sc}$, due to the large-$N$ limit, the Dyson equation for the fermion Green's functions is local in space and is simply given by
\begin{align}
&\Sigma(\tau) = -J^2 G^4(\tau)G^3(-\tau) - K^2 G^2(\tau)G(-\tau), \nn
&G^{-1}(i\omega_n) = i\omega_n-\Sigma(i\omega_n),
\label{DysonNormal}
\end{align}
for both fermion types $a$ and $b$. Defining the energy scaling dimensions $[a]=[b]=1/4$ using the $K$ terms in (\ref{fullH}), we see that $J$ is irrelevant at low energies, but dominates at high energies. This implies a crossover between two distinct IM states around $T\approx K^2/J$: An SYK$_8$ dominant regime with the Green's function given in Eq.~(\ref{GSYK8}) for $T\gg K^2/J$ and an SYK$_4$ dominant regime with the Green's function given in Eq.~(\ref{GSYK4}) for $T\ll K^2/J$. Depending on the strength of the attractive interaction $U$ superconductivity will set in out of IM states with qualitatively different transport. 

\begin{figure}
\includegraphics[width=.48\textwidth]{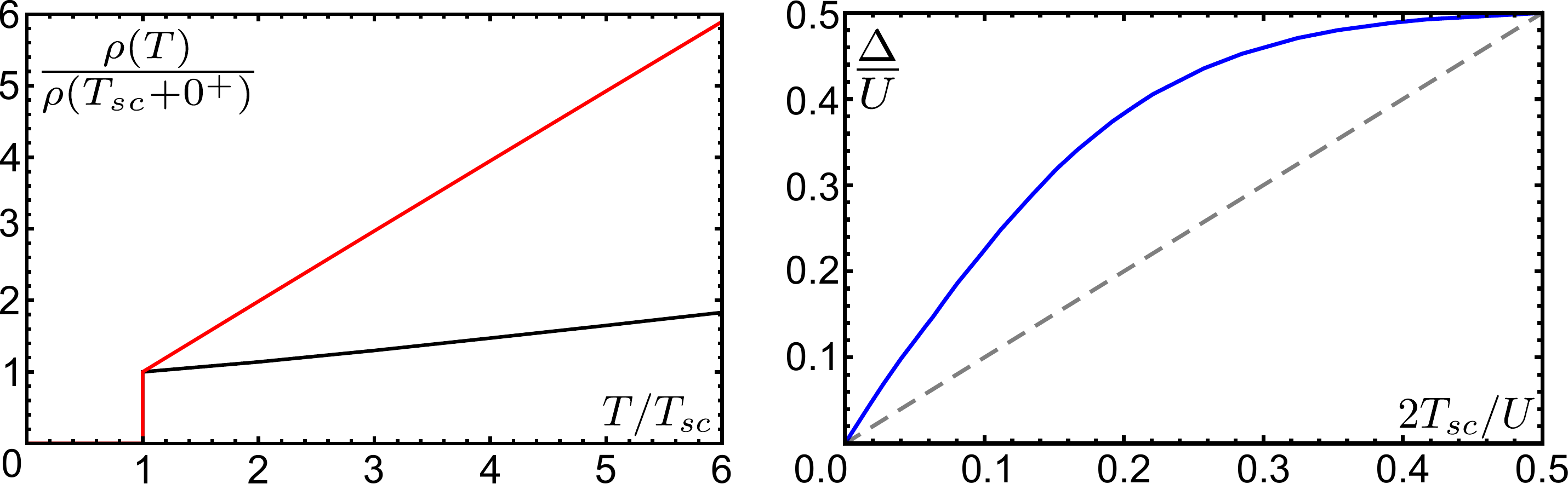}
\caption{{\bf Left panel:} Temperature dependence of the DC resistivity ($\rho(T)/\rho(T_{sc}+0^+)$), vs $T/T_{sc}$ in Model 2. (i) $J\gg K> U$, with a transition from a roughly $T$-independent resistivity to an SC (black), (ii) $J\gg U\gg K$, showing a direct transition from $T$-linear resistivity to an SC (red). The values of the parameters used are (i) $J=100$, $K=20$, $U=13.8584$ and (ii) $J=100$, $U=9.0515$, $K=1$. $T_{sc}=1$ in both cases. {\bf Right panel:} A plot of the normalized physical gap $\Delta/U$ as $T\rightarrow0$ in Model 2, obtained from the spectral function $\mathcal{A}(\omega)$ determined numerically (Supplementary Information), vs the lower bound $\Delta_b=2T_{sc}/U$ (dashed line) for $J=100$, $K=1$, and different values of $U$.}
\label{fig:transportgap2}
\end{figure}

The current operator on the bond indexed by $\langle m n\rangle$, that leads to a conductivity {\it not} suppressed by $1/N$ is
\beq
I_m^{\hat{mn}}= 2i\sum_{i1,..,i4=1}^N (K^{a\langle mn\rangle}_{i1,..i4}a^\dg_{i1m}a^\dg_{i2m}a_{i3n}a_{i4n}+ 
(a\leftrightarrow b)
-\mathrm{H.c.}),
\label{current}
\eeq
(there is another contribution to the current from the $U$ term, but it leads to a contribution to the conductivity that is not extensive in $N$ in the IM). Using this, we then obtain the uniform, disorder averaged, current-current correlator at large-$N$
\beq
\ll \langle I^{\hat{l}} I^{\hat{l}}\rangle({\bf q}=0,\tau) \gg =  4\frac{NK^2}{z}G^2(\tau)G^2(-\tau).
\label{JJ2}
\eeq
For $T\gg K^2/J$, and we can approximate $G$ to be the $\mathrm{SYK}_8$ Green's function (\ref{GSYK8}) to obtain 
\beq
\sigma_{\mathrm{DC}}^J = 2C_8^4\frac{NK^2}{zJT}.
\eeq
Hence this regime is an IM with $T$-linear resistivity. 
However for  $T \ll K^2/J$, the system will cross over to transport controlled by the $\mathrm{SYK}_4$ Green's function (\ref{GSYK4}) with $\sigma_{\mathrm{DC}}^K = N/z$ ({\it i.e.}, a $T$-independent constant). Depending on the relative strength of the attractive interaction $U$ in comparison to the ${\rm SYK}_4$ interaction strength $K$, this $T$-independent resistivity IM may or may not be visible (see Fig.~\ref{fig:transportgap2}).

{\it Superconductivity of Model 2} --
The attractive $U$ term leads to a leading uniform ($\mathbf{q}=0$) $s$-wave pairing instability in the IM phase, which can be seen by considering the renormalization of the $U$ term of (\ref{fullH}) in the pairing channel at different values of the external momentum $\mathbf{q}$, through the standard resummation of pairing bubbles. 
For infinitesimal $U$ with $J\gg K\gg U$, we have a transition from an IM with an approximately $T$-independent resistivity to an SC. In this case the physics of $\rm{SYK}_4$ controls $T_{sc}$ and $T_{sc}$ takes the same form as that in Model 1, given by Eq.~(\ref{Tc42}). A new case of interest is accessible when $J\gg U\gg K$. In this regime the superconducting transition occurs in the temperature range with $T$-linear resistivity and we obtain (Supplementary Information)
\beq
T_{sc} = \frac{C_8^4\Gamma^2(1/4)}{\pi\Gamma^2(3/4)}\frac{U^2}{J}.
\eeq
At $T_{sc}$, we then have a transition from an IM with a linear-in-$T$ resistivity to an $s$-wave SC. 

Now we can investigate implications of the IM normal state on the superconducting state. We find the correlation driven IM normal state affects the superconducting state through an enhancement of the gap ratio as in Model 1. To see this without a BCS limit to benchmark, we consider the limit of vanishing SYK interactions, {\it i.e.}, $U\gg J,K$. In this limit, the paired state becomes entropically unstable above transition temperature of $T_{sc}=U/4$, and the normal state contains featureless free fermions. Further one can find analytically that the zero temperature gap in this limit to be given by $\Delta_b=U/2=2T_{sc}$. Now the implication of IM normal state is apparent  in the numerically obtained value of $\Delta$ (see Fig.~\ref{fig:transportgap2}). $\Delta$ always exceeds the lower bound value of $\Delta_b=2T_{sc}$ (dashed line) in the presence of the $J,K$ interactions. As in Model 1, we find that the superconducting transition becomes first order when the $J,K$ interactions are strong enough to enhance the gap ratio significantly. 

Our models also show coherent superfluid transport despite incoherence driven by the SYK interactions in the normal state. In the SC phase of Model 2, charge transport for $T\ll \Delta$ is controlled by gapless low energy phase fluctuations, whose Hamiltionian is derived by letting $\Delta_{0m} = \Delta_0 e^{i\theta_m}$,
\beq
H_\theta \approx \frac{NU}{z}|\Delta_0|^2\sum_{\langle m,n\rangle} (\theta_m-\theta_n)^2 \rightarrow \frac{NU}{z}|\Delta_0|^2\int_x (\nabla \theta)^2. 
\eeq 
This implies the usual diamagnetic electromagnetic response at frequencies $|\omega|\ll \Delta$ and $T=0$~\cite{Altland2010},
with 
\beq
\lim_{\omega\rightarrow0}\sigma_{ij}(\omega,{\bf q}={\bf 0}) \approx 4\delta_{ij}\frac{NU}{z}\frac{|\Delta_0|^2}{i\omega},
\eeq
and a superfluid phase stiffness that is extensive in $N$, uninhibited by incoherence in the normal state. A similar analysis confirms coherent superfluidity in the superconducting phase of Model 1 (Supplementary Information). 

{\it Conclusion} --
We studied two models exhibiting superconducting transition out of an IM phase with $T$-linear resistivity within the framework of connected SYK ``quantum dots". By having a solvable limit exhibiting this phemonena ubiquitous in correlated systems, we explicitly established implications of a strongly correlated incoherent normal state on superconductivity. The severe electron-electron scattering that destroys coherent quasi-particles and drives $T$-linear resistivity does not inhibit formation of a coherent superconducting state. Instead, the electron-electron scattering leads to dramatic enhancement in the gap ratio, while also driving the superconducting transition first-order.

It is instructive to contrast the gap ratio enhancement seen in our IM-SC transition to that obtained in the standard Eliashberg theory of phonon-mediated superconductivity. Within Eliashberg theory, a relatively gentle deviation of the measured gap ratio from the universal BCS value in elemental superconductors and alloys can be accounted for~\cite{carbotte-RMP1990,PhysRevB.83.064518}. Such enhancement is driven by a suppression of the $T_{sc}$ due to fluctuation effects ignored in the BCS mean-field theory. However, due to the large-$N$ limit, the effects of retardation of the pairing interaction on the Fermion self-energy are suppressed, and the gap equations in (\ref{DysonGap42}) are actually exact. Thus, the enhancement of the gap ratio in our model occurs not due to the suppression of $T_{sc}$ by the thermal fluctuations of the anomalous Green's function, but rather due to the non-quasiparticle nature of the IM Green's functions. It is worth noting that an enhancement of the gap ratio is also seen in holographic models of superconductors~\cite{Hartnoll2008}.

Interestingly, an extreme gap ratio enhancement is widely seen in various correlated-electron superconductors, such as in cuprates and iron-based superconductors~\cite{Hashimoto2014,Hayes2016}. Our work presents the first microscopic mechanism of such an enhancement that is not driven by the suppression of $T_{sc}$ by pairing fluctuations, but rather through the redistribution of spectral weight of an incoherent, non-Fermi liquid normal state.

{\it Acknowledgements} -- We acknowledge useful discussions with James Analytis, Leon Balents, Hong Ding, Steve Kivelson, and Subir Sachdev. This work was initiated during a KITP program supported by NSF Grant No. NSF PHY-1748958. AAP was supported by the NSF Grant PHY-1125915 via a KITP Graduate Fellowship, and by NSF Grant DMR-1664842. E-AK was supported by U.S. Department of Energy, Office of Basic Energy Sciences, Division of Materials Science and Engineering under Award DE-SC0010313.  

\bibliography{sc}

\begin{thebibliography}{27}%
\makeatletter
\providecommand \@ifxundefined [1]{%
 \@ifx{#1\undefined}
}%
\providecommand \@ifnum [1]{%
 \ifnum #1\expandafter \@firstoftwo
 \else \expandafter \@secondoftwo
 \fi
}%
\providecommand \@ifx [1]{%
 \ifx #1\expandafter \@firstoftwo
 \else \expandafter \@secondoftwo
 \fi
}%
\providecommand \natexlab [1]{#1}%
\providecommand \enquote  [1]{``#1''}%
\providecommand \bibnamefont  [1]{#1}%
\providecommand \bibfnamefont [1]{#1}%
\providecommand \citenamefont [1]{#1}%
\providecommand \href@noop [0]{\@secondoftwo}%
\providecommand \href [0]{\begingroup \@sanitize@url \@href}%
\providecommand \@href[1]{\@@startlink{#1}\@@href}%
\providecommand \@@href[1]{\endgroup#1\@@endlink}%
\providecommand \@sanitize@url [0]{\catcode `\\12\catcode `\$12\catcode
  `\&12\catcode `\#12\catcode `\^12\catcode `\_12\catcode `\%12\relax}%
\providecommand \@@startlink[1]{}%
\providecommand \@@endlink[0]{}%
\providecommand \url  [0]{\begingroup\@sanitize@url \@url }%
\providecommand \@url [1]{\endgroup\@href {#1}{\urlprefix }}%
\providecommand \urlprefix  [0]{URL }%
\providecommand \Eprint [0]{\href }%
\providecommand \doibase [0]{http://dx.doi.org/}%
\providecommand \selectlanguage [0]{\@gobble}%
\providecommand \bibinfo  [0]{\@secondoftwo}%
\providecommand \bibfield  [0]{\@secondoftwo}%
\providecommand \translation [1]{[#1]}%
\providecommand \BibitemOpen [0]{}%
\providecommand \bibitemStop [0]{}%
\providecommand \bibitemNoStop [0]{.\EOS\space}%
\providecommand \EOS [0]{\spacefactor3000\relax}%
\providecommand \BibitemShut  [1]{\csname bibitem#1\endcsname}%
\let\auto@bib@innerbib\@empty
\bibitem [{\citenamefont {Varma}\ \emph {et~al.}(1989)\citenamefont {Varma},
  \citenamefont {Littlewood}, \citenamefont {Schmitt-Rink}, \citenamefont
  {Abrahams},\ and\ \citenamefont {Ruckenstein}}]{Varma-PhysRevLett.63.1996}%
  \BibitemOpen
  \bibfield  {author} {\bibinfo {author} {\bibfnamefont {C.~M.}\ \bibnamefont
  {Varma}}, \bibinfo {author} {\bibfnamefont {P.~B.}\ \bibnamefont
  {Littlewood}}, \bibinfo {author} {\bibfnamefont {S.}~\bibnamefont
  {Schmitt-Rink}}, \bibinfo {author} {\bibfnamefont {E.}~\bibnamefont
  {Abrahams}}, \ and\ \bibinfo {author} {\bibfnamefont {A.~E.}\ \bibnamefont
  {Ruckenstein}},\ }\href {\doibase 10.1103/PhysRevLett.63.1996} {\bibfield
  {journal} {\bibinfo  {journal} {Phys. Rev. Lett.}\ }\textbf {\bibinfo
  {volume} {63}},\ \bibinfo {pages} {1996} (\bibinfo {year}
  {1989})}\BibitemShut {NoStop}%
\bibitem [{\citenamefont {Hartnoll}(2014)}]{Hartnoll-NP14}%
  \BibitemOpen
  \bibfield  {author} {\bibinfo {author} {\bibfnamefont {S.~A.}\ \bibnamefont
  {Hartnoll}},\ }\href {http://dx.doi.org/10.1038/nphys3174} {\bibfield
  {journal} {\bibinfo  {journal} {Nature Physics}\ }\textbf {\bibinfo {volume}
  {11}},\ \bibinfo {pages} {54 EP } (\bibinfo {year} {2014})}\BibitemShut
  {NoStop}%
\bibitem [{\citenamefont {Hartnoll}\ and\ \citenamefont
  {Karch}(2015)}]{Hartnoll-Karch-PhysRevB.91.155126}%
  \BibitemOpen
  \bibfield  {author} {\bibinfo {author} {\bibfnamefont {S.~A.}\ \bibnamefont
  {Hartnoll}}\ and\ \bibinfo {author} {\bibfnamefont {A.}~\bibnamefont
  {Karch}},\ }\href {\doibase 10.1103/PhysRevB.91.155126} {\bibfield  {journal}
  {\bibinfo  {journal} {Phys. Rev. B}\ }\textbf {\bibinfo {volume} {91}},\
  \bibinfo {pages} {155126} (\bibinfo {year} {2015})}\BibitemShut {NoStop}%
\bibitem [{\citenamefont {Bruin}\ \emph {et~al.}(2013)\citenamefont {Bruin},
  \citenamefont {Sakai}, \citenamefont {Perry},\ and\ \citenamefont
  {Mackenzie}}]{Bruin804}%
  \BibitemOpen
  \bibfield  {author} {\bibinfo {author} {\bibfnamefont {J.~A.~N.}\
  \bibnamefont {Bruin}}, \bibinfo {author} {\bibfnamefont {H.}~\bibnamefont
  {Sakai}}, \bibinfo {author} {\bibfnamefont {R.~S.}\ \bibnamefont {Perry}}, \
  and\ \bibinfo {author} {\bibfnamefont {A.~P.}\ \bibnamefont {Mackenzie}},\
  }\href {\doibase 10.1126/science.1227612} {\bibfield  {journal} {\bibinfo
  {journal} {Science}\ }\textbf {\bibinfo {volume} {339}},\ \bibinfo {pages}
  {804} (\bibinfo {year} {2013})}\BibitemShut {NoStop}%
\bibitem [{\citenamefont {Emery}\ and\ \citenamefont
  {Kivelson}(1995{\natexlab{a}})}]{emery-kivelson1995-nature}%
  \BibitemOpen
  \bibfield  {author} {\bibinfo {author} {\bibfnamefont {V.~J.}\ \bibnamefont
  {Emery}}\ and\ \bibinfo {author} {\bibfnamefont {S.~A.}\ \bibnamefont
  {Kivelson}},\ }\href {http://dx.doi.org/10.1038/374434a0} {\bibfield
  {journal} {\bibinfo  {journal} {Nature}\ }\textbf {\bibinfo {volume} {374}},\
  \bibinfo {pages} {434 EP } (\bibinfo {year}
  {1995}{\natexlab{a}})}\BibitemShut {NoStop}%
\bibitem [{\citenamefont {Emery}\ and\ \citenamefont
  {Kivelson}(1995{\natexlab{b}})}]{Emery-Kivelson95PRL-PhysRevLett.74.3253}%
  \BibitemOpen
  \bibfield  {author} {\bibinfo {author} {\bibfnamefont {V.~J.}\ \bibnamefont
  {Emery}}\ and\ \bibinfo {author} {\bibfnamefont {S.~A.}\ \bibnamefont
  {Kivelson}},\ }\href {\doibase 10.1103/PhysRevLett.74.3253} {\bibfield
  {journal} {\bibinfo  {journal} {Phys. Rev. Lett.}\ }\textbf {\bibinfo
  {volume} {74}},\ \bibinfo {pages} {3253} (\bibinfo {year}
  {1995}{\natexlab{b}})}\BibitemShut {NoStop}%
\bibitem [{\citenamefont {Zaanen}(2004)}]{Zaanen-Plankian}%
  \BibitemOpen
  \bibfield  {author} {\bibinfo {author} {\bibfnamefont {J.}~\bibnamefont
  {Zaanen}},\ }\href {http://dx.doi.org/10.1038/430512a} {\bibfield  {journal}
  {\bibinfo  {journal} {Nature}\ }\textbf {\bibinfo {volume} {430}},\ \bibinfo
  {pages} {512 EP } (\bibinfo {year} {2004})}\BibitemShut {NoStop}%
\bibitem [{\citenamefont {Gu}\ \emph {et~al.}(2017)\citenamefont {Gu},
  \citenamefont {Qi},\ and\ \citenamefont {Stanford}}]{Gu2017}%
  \BibitemOpen
  \bibfield  {author} {\bibinfo {author} {\bibfnamefont {Y.}~\bibnamefont
  {Gu}}, \bibinfo {author} {\bibfnamefont {X.-L.}\ \bibnamefont {Qi}}, \ and\
  \bibinfo {author} {\bibfnamefont {D.}~\bibnamefont {Stanford}},\ }\href
  {\doibase 10.1007/JHEP05(2017)125} {\bibfield  {journal} {\bibinfo  {journal}
  {Journal of High Energy Physics}\ }\textbf {\bibinfo {volume} {2017}},\
  \bibinfo {pages} {125} (\bibinfo {year} {2017})}\BibitemShut {NoStop}%
\bibitem [{\citenamefont {{Davison}}\ \emph {et~al.}(2017)\citenamefont
  {{Davison}}, \citenamefont {{Fu}}, \citenamefont {{Georges}}, \citenamefont
  {{Gu}}, \citenamefont {{Jensen}},\ and\ \citenamefont
  {{Sachdev}}}]{Sachdev2017}%
  \BibitemOpen
  \bibfield  {author} {\bibinfo {author} {\bibfnamefont {R.~A.}\ \bibnamefont
  {{Davison}}}, \bibinfo {author} {\bibfnamefont {W.}~\bibnamefont {{Fu}}},
  \bibinfo {author} {\bibfnamefont {A.}~\bibnamefont {{Georges}}}, \bibinfo
  {author} {\bibfnamefont {Y.}~\bibnamefont {{Gu}}}, \bibinfo {author}
  {\bibfnamefont {K.}~\bibnamefont {{Jensen}}}, \ and\ \bibinfo {author}
  {\bibfnamefont {S.}~\bibnamefont {{Sachdev}}},\ }\href {\doibase
  10.1103/PhysRevB.95.155131} {\bibfield  {journal} {\bibinfo  {journal} {Phys.
  Rev. B}\ }\textbf {\bibinfo {volume} {95}},\ \bibinfo {eid} {155131}
  (\bibinfo {year} {2017})},\ \Eprint {http://arxiv.org/abs/1612.00849}
  {arXiv:1612.00849 [cond-mat.str-el]} \BibitemShut {NoStop}%
\bibitem [{\citenamefont {Song}\ \emph {et~al.}(2017)\citenamefont {Song},
  \citenamefont {Jian},\ and\ \citenamefont {Balents}}]{Balents2017}%
  \BibitemOpen
  \bibfield  {author} {\bibinfo {author} {\bibfnamefont {X.-Y.}\ \bibnamefont
  {Song}}, \bibinfo {author} {\bibfnamefont {C.-M.}\ \bibnamefont {Jian}}, \
  and\ \bibinfo {author} {\bibfnamefont {L.}~\bibnamefont {Balents}},\ }\href
  {\doibase 10.1103/PhysRevLett.119.216601} {\bibfield  {journal} {\bibinfo
  {journal} {Phys. Rev. Lett.}\ }\textbf {\bibinfo {volume} {119}},\ \bibinfo
  {pages} {216601} (\bibinfo {year} {2017})}\BibitemShut {NoStop}%
\bibitem [{\citenamefont {Zhang}(2017)}]{Zhang2017}%
  \BibitemOpen
  \bibfield  {author} {\bibinfo {author} {\bibfnamefont {P.}~\bibnamefont
  {Zhang}},\ }\href {\doibase 10.1103/PhysRevB.96.205138} {\bibfield  {journal}
  {\bibinfo  {journal} {Phys. Rev. B}\ }\textbf {\bibinfo {volume} {96}},\
  \bibinfo {pages} {205138} (\bibinfo {year} {2017})}\BibitemShut {NoStop}%
\bibitem [{\citenamefont {Patel}\ \emph {et~al.}(2018)\citenamefont {Patel},
  \citenamefont {McGreevy}, \citenamefont {Arovas},\ and\ \citenamefont
  {Sachdev}}]{Patel2017}%
  \BibitemOpen
  \bibfield  {author} {\bibinfo {author} {\bibfnamefont {A.~A.}\ \bibnamefont
  {Patel}}, \bibinfo {author} {\bibfnamefont {J.}~\bibnamefont {McGreevy}},
  \bibinfo {author} {\bibfnamefont {D.~P.}\ \bibnamefont {Arovas}}, \ and\
  \bibinfo {author} {\bibfnamefont {S.}~\bibnamefont {Sachdev}},\ }\href
  {\doibase 10.1103/PhysRevX.8.021049} {\bibfield  {journal} {\bibinfo
  {journal} {Phys. Rev. X}\ }\textbf {\bibinfo {volume} {8}},\ \bibinfo {pages}
  {021049} (\bibinfo {year} {2018})}\BibitemShut {NoStop}%
\bibitem [{\citenamefont {Chowdhury}\ \emph {et~al.}(2018)\citenamefont
  {Chowdhury}, \citenamefont {Werman}, \citenamefont {Berg},\ and\
  \citenamefont {Senthil}}]{Chowdhury2018}%
  \BibitemOpen
  \bibfield  {author} {\bibinfo {author} {\bibfnamefont {D.}~\bibnamefont
  {Chowdhury}}, \bibinfo {author} {\bibfnamefont {Y.}~\bibnamefont {Werman}},
  \bibinfo {author} {\bibfnamefont {E.}~\bibnamefont {Berg}}, \ and\ \bibinfo
  {author} {\bibfnamefont {T.}~\bibnamefont {Senthil}},\ }\href {\doibase
  10.1103/PhysRevX.8.031024} {\bibfield  {journal} {\bibinfo  {journal} {Phys.
  Rev. X}\ }\textbf {\bibinfo {volume} {8}},\ \bibinfo {pages} {031024}
  (\bibinfo {year} {2018})}\BibitemShut {NoStop}%
\bibitem [{\citenamefont {{Wu}}\ \emph {et~al.}(2018)\citenamefont {{Wu}},
  \citenamefont {{Chen}}, \citenamefont {{Jian}}, \citenamefont {{You}},\ and\
  \citenamefont {{Xu}}}]{Wu2018}%
  \BibitemOpen
  \bibfield  {author} {\bibinfo {author} {\bibfnamefont {X.}~\bibnamefont
  {{Wu}}}, \bibinfo {author} {\bibfnamefont {X.}~\bibnamefont {{Chen}}},
  \bibinfo {author} {\bibfnamefont {C.-M.}\ \bibnamefont {{Jian}}}, \bibinfo
  {author} {\bibfnamefont {Y.-Z.}\ \bibnamefont {{You}}}, \ and\ \bibinfo
  {author} {\bibfnamefont {C.}~\bibnamefont {{Xu}}},\ }\href@noop {} {\bibfield
   {journal} {\bibinfo  {journal} {ArXiv e-prints}\ } (\bibinfo {year}
  {2018})},\ \Eprint {http://arxiv.org/abs/1802.04293} {arXiv:1802.04293
  [cond-mat.str-el]} \BibitemShut {NoStop}%
\bibitem [{\citenamefont {Sachdev}\ and\ \citenamefont
  {Ye}(1993)}]{Sachdev-Ye}%
  \BibitemOpen
  \bibfield  {author} {\bibinfo {author} {\bibfnamefont {S.}~\bibnamefont
  {Sachdev}}\ and\ \bibinfo {author} {\bibfnamefont {J.}~\bibnamefont {Ye}},\
  }\href {\doibase 10.1103/PhysRevLett.70.3339} {\bibfield  {journal} {\bibinfo
   {journal} {Phys. Rev. Lett.}\ }\textbf {\bibinfo {volume} {70}},\ \bibinfo
  {pages} {3339} (\bibinfo {year} {1993})}\BibitemShut {NoStop}%
\bibitem [{\citenamefont {{Kitaev}}(2015)}]{Kitaev}%
  \BibitemOpen
  \bibfield  {author} {\bibinfo {author} {\bibfnamefont {A.~Y.}\ \bibnamefont
  {{Kitaev}}},\ }\href {http://online.kitp.ucsb.edu/online/entangled15/}
  {\bibfield  {journal} {\bibinfo  {journal} {Entanglement in
  Strongly-Correlated Quantum Matter}\ } (\bibinfo {year} {2015})}\BibitemShut
  {NoStop}%
\bibitem [{\citenamefont {Sachdev}(2015)}]{Sachdev2015}%
  \BibitemOpen
  \bibfield  {author} {\bibinfo {author} {\bibfnamefont {S.}~\bibnamefont
  {Sachdev}},\ }\href {\doibase 10.1103/PhysRevX.5.041025} {\bibfield
  {journal} {\bibinfo  {journal} {Phys. Rev. X}\ }\textbf {\bibinfo {volume}
  {5}},\ \bibinfo {pages} {041025} (\bibinfo {year} {2015})}\BibitemShut
  {NoStop}%
\bibitem [{\citenamefont {Bardeen}\ \emph {et~al.}(1957)\citenamefont
  {Bardeen}, \citenamefont {Cooper},\ and\ \citenamefont {Schrieffer}}]{BCS}%
  \BibitemOpen
  \bibfield  {author} {\bibinfo {author} {\bibfnamefont {J.}~\bibnamefont
  {Bardeen}}, \bibinfo {author} {\bibfnamefont {L.~N.}\ \bibnamefont {Cooper}},
  \ and\ \bibinfo {author} {\bibfnamefont {J.~R.}\ \bibnamefont {Schrieffer}},\
  }\href {\doibase 10.1103/PhysRev.108.1175} {\bibfield  {journal} {\bibinfo
  {journal} {Phys. Rev.}\ }\textbf {\bibinfo {volume} {108}},\ \bibinfo {pages}
  {1175} (\bibinfo {year} {1957})}\BibitemShut {NoStop}%
\bibitem [{\citenamefont {Chubukov}\ \emph {et~al.}(2003)\citenamefont
  {Chubukov}, \citenamefont {Finkel'stein}, \citenamefont {Haslinger},\ and\
  \citenamefont {Morr}}]{Chubukov2003}%
  \BibitemOpen
  \bibfield  {author} {\bibinfo {author} {\bibfnamefont {A.~V.}\ \bibnamefont
  {Chubukov}}, \bibinfo {author} {\bibfnamefont {A.~M.}\ \bibnamefont
  {Finkel'stein}}, \bibinfo {author} {\bibfnamefont {R.}~\bibnamefont
  {Haslinger}}, \ and\ \bibinfo {author} {\bibfnamefont {D.~K.}\ \bibnamefont
  {Morr}},\ }\href {\doibase 10.1103/PhysRevLett.90.077002} {\bibfield
  {journal} {\bibinfo  {journal} {Phys. Rev. Lett.}\ }\textbf {\bibinfo
  {volume} {90}},\ \bibinfo {pages} {077002} (\bibinfo {year}
  {2003})}\BibitemShut {NoStop}%
\bibitem [{\citenamefont {Giannakis}\ \emph {et~al.}(2004)\citenamefont
  {Giannakis}, \citenamefont {Hou}, \citenamefont {Ren},\ and\ \citenamefont
  {Rischke}}]{Giannakis2004}%
  \BibitemOpen
  \bibfield  {author} {\bibinfo {author} {\bibfnamefont {I.}~\bibnamefont
  {Giannakis}}, \bibinfo {author} {\bibfnamefont {D.}~\bibnamefont {Hou}},
  \bibinfo {author} {\bibfnamefont {H.-c.}\ \bibnamefont {Ren}}, \ and\
  \bibinfo {author} {\bibfnamefont {D.~H.}\ \bibnamefont {Rischke}},\ }\href
  {\doibase 10.1103/PhysRevLett.93.232301} {\bibfield  {journal} {\bibinfo
  {journal} {Phys. Rev. Lett.}\ }\textbf {\bibinfo {volume} {93}},\ \bibinfo
  {pages} {232301} (\bibinfo {year} {2004})}\BibitemShut {NoStop}%
\bibitem [{\citenamefont {Altland}\ and\ \citenamefont
  {Simons}(2010)}]{Altland2010}%
  \BibitemOpen
  \bibfield  {author} {\bibinfo {author} {\bibfnamefont {A.}~\bibnamefont
  {Altland}}\ and\ \bibinfo {author} {\bibfnamefont {B.~D.}\ \bibnamefont
  {Simons}},\ }\href
  {http://www.cambridge.org/gb/academic/subjects/physics/condensed-matter-physics-nanoscience-and-mesoscopic-physics/condensed-matter-field-theory-2nd-edition?format=HB&isbn=9780521769754#oMPp6LOXAPexXdyb.97}
  {\emph {\bibinfo {title} {Condensed matter field theory}}}\ (\bibinfo
  {publisher} {Cambridge University Press},\ \bibinfo {year}
  {2010})\BibitemShut {NoStop}%
\bibitem [{\citenamefont {Carbotte}(1990)}]{carbotte-RMP1990}%
  \BibitemOpen
  \bibfield  {author} {\bibinfo {author} {\bibfnamefont {J.~P.}\ \bibnamefont
  {Carbotte}},\ }\href {\doibase 10.1103/RevModPhys.62.1027} {\bibfield
  {journal} {\bibinfo  {journal} {Rev. Mod. Phys.}\ }\textbf {\bibinfo {volume}
  {62}},\ \bibinfo {pages} {1027} (\bibinfo {year} {1990})}\BibitemShut
  {NoStop}%
\bibitem [{\citenamefont {Dhokarh}\ and\ \citenamefont
  {Chubukov}(2011)}]{PhysRevB.83.064518}%
  \BibitemOpen
  \bibfield  {author} {\bibinfo {author} {\bibfnamefont {D.}~\bibnamefont
  {Dhokarh}}\ and\ \bibinfo {author} {\bibfnamefont {A.~V.}\ \bibnamefont
  {Chubukov}},\ }\href {\doibase 10.1103/PhysRevB.83.064518} {\bibfield
  {journal} {\bibinfo  {journal} {Phys. Rev. B}\ }\textbf {\bibinfo {volume}
  {83}},\ \bibinfo {pages} {064518} (\bibinfo {year} {2011})}\BibitemShut
  {NoStop}%
\bibitem [{\citenamefont {Hartnoll}\ \emph {et~al.}(2008)\citenamefont
  {Hartnoll}, \citenamefont {Herzog},\ and\ \citenamefont
  {Horowitz}}]{Hartnoll2008}%
  \BibitemOpen
  \bibfield  {author} {\bibinfo {author} {\bibfnamefont {S.~A.}\ \bibnamefont
  {Hartnoll}}, \bibinfo {author} {\bibfnamefont {C.~P.}\ \bibnamefont
  {Herzog}}, \ and\ \bibinfo {author} {\bibfnamefont {G.~T.}\ \bibnamefont
  {Horowitz}},\ }\href {\doibase 10.1103/PhysRevLett.101.031601} {\bibfield
  {journal} {\bibinfo  {journal} {Phys. Rev. Lett.}\ }\textbf {\bibinfo
  {volume} {101}},\ \bibinfo {pages} {031601} (\bibinfo {year}
  {2008})}\BibitemShut {NoStop}%
\bibitem [{\citenamefont {Hashimoto}\ \emph {et~al.}(2014)\citenamefont
  {Hashimoto}, \citenamefont {Vishik}, \citenamefont {He}, \citenamefont
  {Devereaux},\ and\ \citenamefont {Shen}}]{Hashimoto2014}%
  \BibitemOpen
  \bibfield  {author} {\bibinfo {author} {\bibfnamefont {M.}~\bibnamefont
  {Hashimoto}}, \bibinfo {author} {\bibfnamefont {I.~M.}\ \bibnamefont
  {Vishik}}, \bibinfo {author} {\bibfnamefont {R.-H.}\ \bibnamefont {He}},
  \bibinfo {author} {\bibfnamefont {T.~P.}\ \bibnamefont {Devereaux}}, \ and\
  \bibinfo {author} {\bibfnamefont {Z.-X.}\ \bibnamefont {Shen}},\ }\href
  {http://dx.doi.org/10.1038/nphys3009} {\bibfield  {journal} {\bibinfo
  {journal} {Nature Physics}\ }\textbf {\bibinfo {volume} {10}},\ \bibinfo
  {pages} {483 EP } (\bibinfo {year} {2014})}\BibitemShut {NoStop}%
\bibitem [{\citenamefont {Hayes}\ \emph {et~al.}(2016)\citenamefont {Hayes},
  \citenamefont {McDonald}, \citenamefont {Breznay}, \citenamefont {Helm},
  \citenamefont {Moll}, \citenamefont {Wartenbe}, \citenamefont {Shekhter},\
  and\ \citenamefont {Analytis}}]{Hayes2016}%
  \BibitemOpen
  \bibfield  {author} {\bibinfo {author} {\bibfnamefont {I.~M.}\ \bibnamefont
  {Hayes}}, \bibinfo {author} {\bibfnamefont {R.~D.}\ \bibnamefont {McDonald}},
  \bibinfo {author} {\bibfnamefont {N.~P.}\ \bibnamefont {Breznay}}, \bibinfo
  {author} {\bibfnamefont {T.}~\bibnamefont {Helm}}, \bibinfo {author}
  {\bibfnamefont {P.~J.~W.}\ \bibnamefont {Moll}}, \bibinfo {author}
  {\bibfnamefont {M.}~\bibnamefont {Wartenbe}}, \bibinfo {author}
  {\bibfnamefont {A.}~\bibnamefont {Shekhter}}, \ and\ \bibinfo {author}
  {\bibfnamefont {J.~G.}\ \bibnamefont {Analytis}},\ }\href
  {http://dx.doi.org/10.1038/nphys3773} {\bibfield  {journal} {\bibinfo
  {journal} {Nature Physics}\ }\textbf {\bibinfo {volume} {12}},\ \bibinfo
  {pages} {916 EP } (\bibinfo {year} {2016})}\BibitemShut {NoStop}%
\bibitem [{\citenamefont {Kamenev}(2011)}]{Kamenev2011}%
  \BibitemOpen
  \bibfield  {author} {\bibinfo {author} {\bibfnamefont {A.}~\bibnamefont
  {Kamenev}},\ }\href
  {http://www.cambridge.org/us/academic/subjects/physics/condensed-matter-physics-nanoscience-and-mesoscopic-physics/field-theory-non-equilibrium-systems?format=HB&isbn=9780521760829#3WyIuCiHhbKA00MR.97}
  {\emph {\bibinfo {title} {Field theory of non-equilibrium systems}}}\
  (\bibinfo  {publisher} {Cambridge University Press},\ \bibinfo {year}
  {2011})\BibitemShut {NoStop}%
\end{thebibliography}%

\begin{widetext}

\appendix

\section{Derivation of gap equations}

We derive the combined Dyson and gap equations from the large-$N$ saddle point action. For Model 1, the disorder averaged action can be written as~\cite{Sachdev2015,Sachdev2017,Balents2017,Patel2017}
\begin{align}
&S = \int_0^\beta d\tau \Bigg[\sum_{m,i}\left(a^\dg_{im}(\partial_\tau-\mu)a_{im}+(a\leftrightarrow b)\right) -t\sum_{\langle m,n\rangle,i}\left(a^\dg_{im}a_{in}+(a\leftrightarrow b)+\mathrm{H.c.}\right)\Bigg] \nn
&-N\frac{K^2}{4}\int_0^\beta d\tau d\tau^\prime \sum_m\left(\mathcal{G}_{am}^2(\tau,\tau^\prime)\mathcal{G}_{am}^2(\tau^\prime,\tau)+(a\leftrightarrow b)\right) \nn
&-N\int_0^\beta d\tau d\tau^\prime \sum_m \Bigg[\Sigma_{am}(\tau,\tau^\prime)\Bigg(\mathcal{G}_{am}(\tau^\prime,\tau) +\frac{1}{N}\sum_i a^\dg_{im}(\tau)a_{im}(\tau^\prime)\Bigg)+(a\leftrightarrow b)\Bigg] \nn
&-NU\int_0^\beta d\tau \sum_m \Delta_{0m}^\ast(\tau)\Delta_{0m}(\tau) -N\int_0^\beta\sum_m \Bigg[\Xi_m(\tau)\left(\Delta_{0m}(\tau)-\frac{1}{N}\sum_i a_{im}(\tau)b_{im}(\tau)\right) +\mathrm{H.c}\Bigg].
\label{S1}
\end{align}
In the large-$N$ limit, the Lagrange multiplier fields $\Sigma_{\alpha m}(\tau,\tau^\prime)$ and $\Xi_m(\tau),\Xi^\ast_m(\tau)$ enforce the definitions of $\mathcal{G}$ and $\Delta^\ast_0,\Delta_0$ at each site $m$. After integrating out the fermions, varying the action with respect to $\mathcal{G}_{\alpha m}$ and $\Delta_{0m},\Delta^\ast_{0m}$ yields
\begin{align}
&\Sigma_{\alpha m}(\tau,\tau^\prime) = -K^2\mathcal{G}_{\alpha m}^2(\tau,\tau^\prime)\mathcal{G}_{\alpha m}(\tau^\prime,\tau), \nn
&\Xi_m(\tau)=-U\Delta^\ast_{0m}(\tau),~~\Xi_m^\ast(\tau)=-U\Delta_{0m}(\tau).
\label{SigmaXi}
\end{align}
For the saddle point, we look for a uniform solution in which $\Sigma,\mathcal{G},\Xi,\Delta_0$ are independent of $m$ and $\alpha\equiv a,b$, and where $\Xi$ and $\Delta_0$ are constant in time. Then, additionally varying the action with the fermions integrated out with respect to $\Xi,\Xi^\ast$ and $\Sigma(\tau,\tau^\prime)$ and applying (\ref{SigmaXi}) generates the set of equations in (\ref{DysonGap42}). 

The result (\ref{Tc42}) for $T_{sc}$ in the limit of small bandwidth may then be derived from (\ref{DysonGap42}) by sending $\Delta_0\rightarrow0$ and ignoring the dispersion $\xi_k$. We get
\beq
\int_{K^{-1}}^{T_{sc}^{-1}-K^{-1}} d\tau~G^2(\tau) = \frac{1}{U}.
\eeq
Using the IM Green's function (\ref{GSYK4}) appropriate to this limit, we obtain (\ref{Tc42}).

After integrating out the fermions and taking the saddle point of (\ref{S1}), fluctuations in the pairing order parameter $\Delta_{0m}(\tau)=\Delta_{0}e^{i\theta_m(\tau)}$ only affect the fermion determinant term. Then, the effective action for the long-wavelength condensate phase fluctuations $\theta_m(\tau)$ can be derived following the standard procedure in Ref.~\cite{Altland2010}, yielding a coefficient of $(\nabla\theta)^2$ and hence a superfluid stiffness that is extensive in $N$. Due to the large-$N$ limit, this implies that phase coherence for $T<T_{sc}$ is established for any nonzero values of $t$ and $\Delta_0$.

A procedure similar to the one described in this section can be applied to derive the gap equations for Model 2.

\section{Superconducting transition energetics}

In order to study the details of the IM-SC transition, we compute the free energy density at the large-$N$ saddle point in Model 1
\begin{align}
&\frac{\mathcal{F}}{N}= T\sum_{\omega_n}\int\frac{d^dk}{(2\pi)^d}\ln\left[\frac{\omega_n^2+\xi_k^2+U^2\Delta_0^2}{|G(i\omega_n,k)|^{-2}+U^2\Delta_0^2}\right] - T\int\frac{d^dk}{(2\pi)^d} \ln\left[\left(1+e^{\sqrt{U^2\Delta_0^2+\xi_k^2}/T}\right)\left(1+e^{-\sqrt{U^2\Delta_0^2+\xi_k^2}/T}\right)\right] \nn
&-\frac{3}{2}T\sum_{\omega_n}\Sigma(i\omega_n)\mathcal{G}(i\omega_n) + U\Delta_0^2.
\end{align}
To analyze this as a function of the order parameter $\Delta_0$, we determine $G,\Sigma,\mathcal{G}$ as functions of $\Delta_0$ using (\ref{DysonGap42}), but ignore its last line that determines $\Delta_0$ self-consistently. A plot of $\mathcal{F}/N$ vs $\Delta_0$ for $T\gtrsim T_{sc}$ and $T\lesssim T_{sc}$ is shown in Fig.~\ref{fig:energetics}.

\begin{figure}[b]
\includegraphics[width=.47\textwidth]{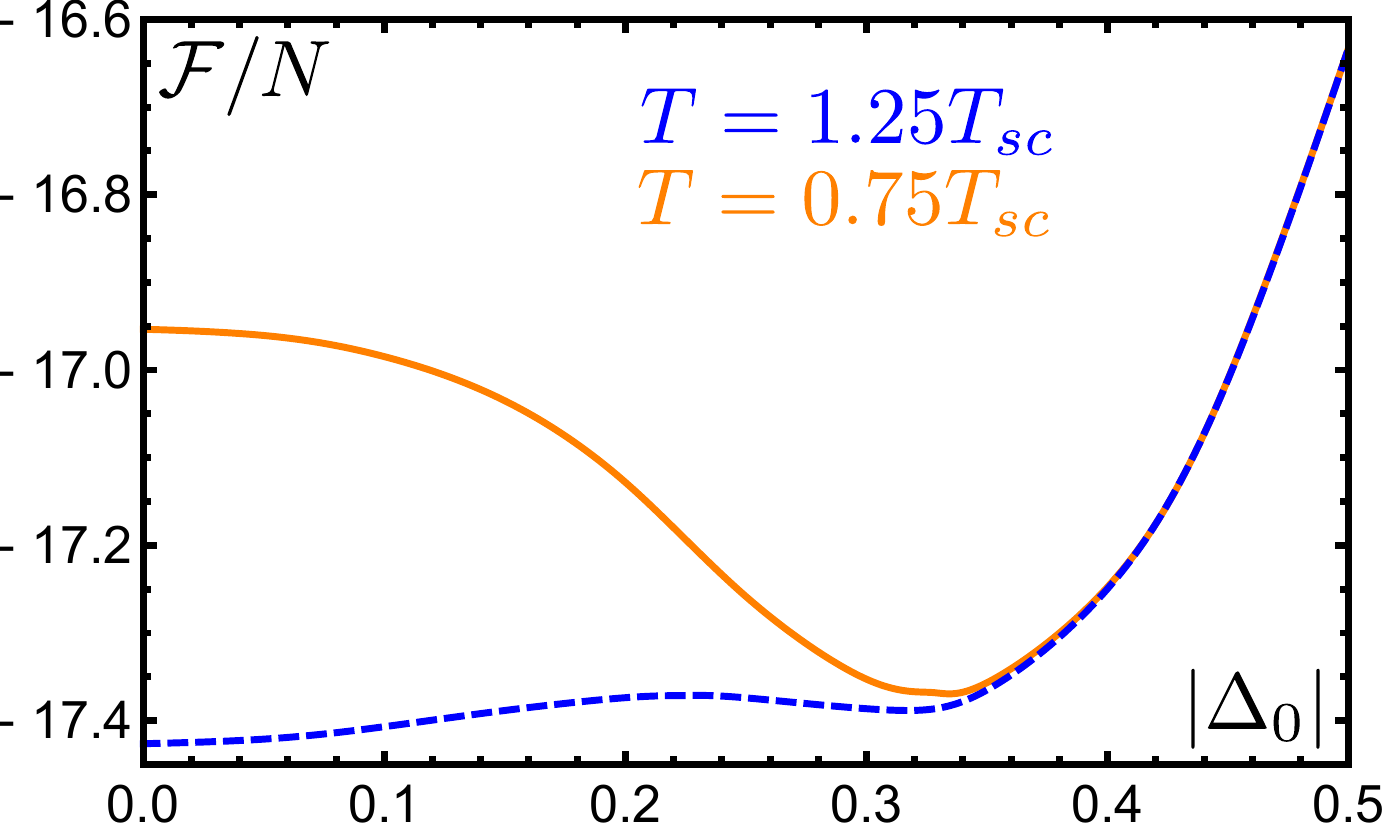}
\caption{A plot of the free energy density in the IM/SC regime in Model 1, as a function of the superconducting order parameter $\Delta_0$. The values of parameters used are $K=100$, $\Lambda=1$, $U=41.34$, so that $T_{sc}=1$.}
\label{fig:energetics}
\end{figure}

The free energy functional is qualitatively similar to that of $\phi^6$ theory, with $\mathcal{F}\sim N(c_2 |\Delta_0|^2 - c_4 |\Delta_0|^4 + c_6 |\Delta_0|^6$), where $c_2\propto(T-T_{sc})$. Thus, as $T$ is lowered below $T_{sc}$ there is a first-order transition from a state with zero gap to a state with a finite gap. The free energies of the gapped and gapless states actually cross each other at a temperature $T_{sc}^{(1)}$ which is slightly larger than $T_{sc}$, but the large-$N$ limit strongly suppresses tunneling from one local minimum to another: the tunneling rate goes as $\sim e^{-N}$ in the WKB approximation.  

A similar effect is also seen for Model 2.

\section{Real-time Dyson equations}

{\it Superconductors}-- 
The real-time Green's functions for the gapped SC phase of Model 1 can be obtained by solving the Dyson equation~(\ref{DysonGap42}) on the Keldysh contour. It is given by
\begin{align}
&\mathcal{G}_R(\omega) = \int \frac{d^dk}{(2\pi)^d}\frac{G_R(\omega,k)}{1+U^2\Delta_0^2G_R(\omega,k)G_R^\ast(-\omega,k)}, \nn
&\mathcal{G}_K(\omega) = 2i\mathrm{Im}[\mathcal{G}_R(\omega)]\tanh(\omega/(2T)), \nn
&\Sigma_R(t) = -\frac{K^2}{8}\theta(t)\Big[(\mathcal{G}_K(t)+\mathcal{G}_R(t))^2(\mathcal{G}^\ast_K(t) -\mathcal{G}^\ast_R(t))-(\mathcal{G}_K(t)-\mathcal{G}_R(t))^2(\mathcal{G}^\ast_K(t)+\mathcal{G}^\ast_R(t))\Big], \nn
&\Sigma_K(t>0) = -\frac{K^2}{8}\Big[(\mathcal{G}_K(t)+\mathcal{G}_R(t))^2(\mathcal{G}^\ast_K(t) -\mathcal{G}^\ast_R(t))+(\mathcal{G}_K(t)-\mathcal{G}_R(t))^2(\mathcal{G}^\ast_K(t)+\mathcal{G}^\ast_R(t))\Big], \nn
&\Sigma_K(t<0) = -\Sigma_K^\ast(-t>0), \nn
&G_R(\omega,k) = 2\tanh(\omega/(2T))\left[2\tanh(\omega/(2T))(\omega-\xi_k+i0^+-\mathrm{Re}[\Sigma_R(\omega)])-\Sigma_K(\omega)\right]^{-1},
\label{DysonKeldysh}
\end{align}
where we exploited the standard simplifications for a system in equilibrium~\cite{Kamenev2011}.

The order parameter $\Delta_0$ is first determined from the solution of the imaginary-time equations (\ref{DysonGap42}), and inserted into the real-time equations, which are then solved iteratively, determining $G_R(\omega)$ and $\mathcal{G}_R(\omega)$. A plot of $\mathcal{A}(\omega,\{k:\xi_k=0\})$ is shown in Fig.~\ref{fig:gapspectral}, clearly showing peaks at the physical gap $\omega=\pm\Delta$. The same strategy can also be applied to Model 2.

\begin{figure}[b]
\includegraphics[width=.47\textwidth]{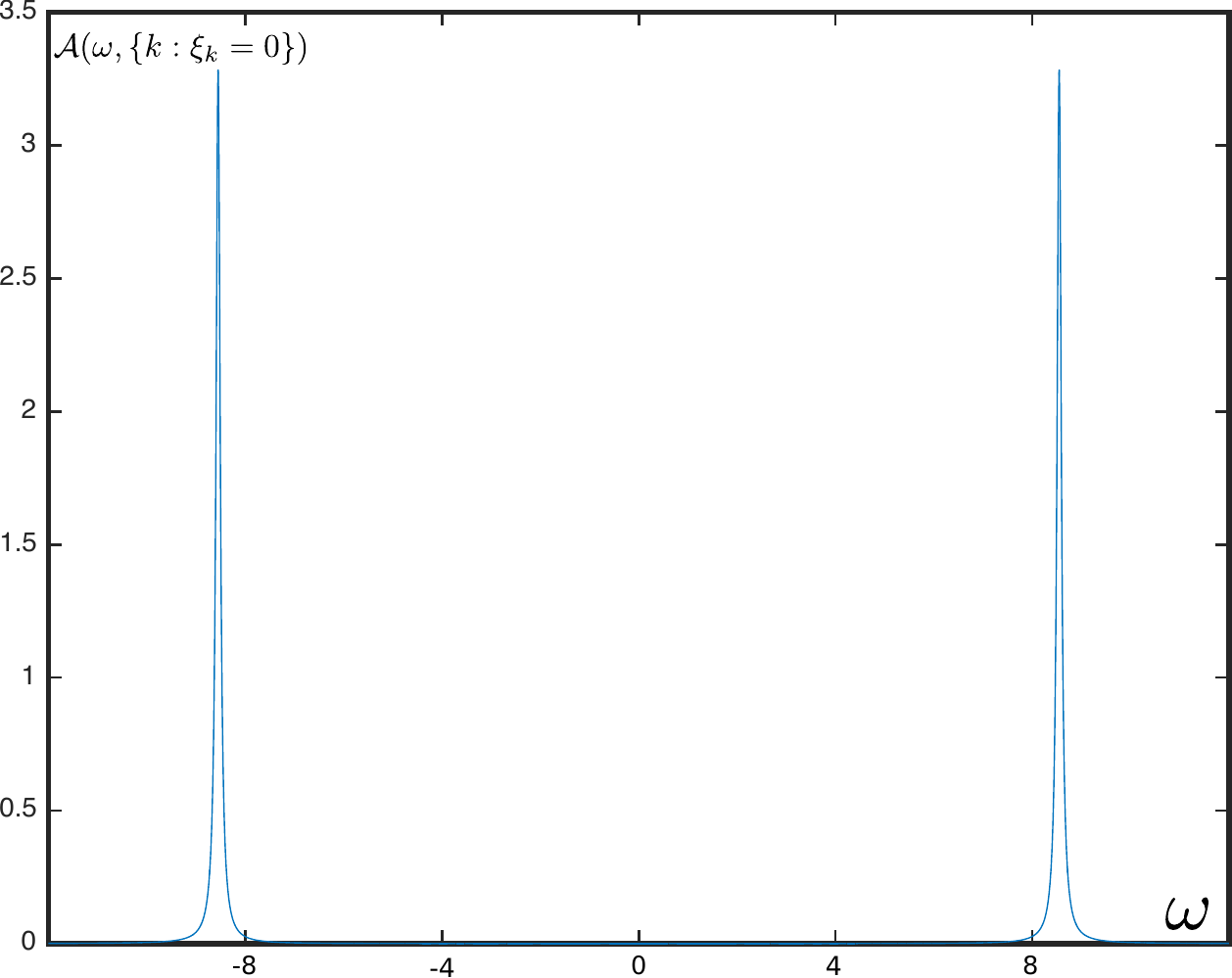}
\caption{A plot of the spectral function $\mathcal{A}(\omega,\{k:\xi_k=0\})$ in the SC phase at of Model 1 as $T\rightarrow0$, for values of parameters such that the SC arises out of the IM. The spectral function is strongly peaked at the physical gap $\omega=\pm \Delta$. The values of the parameters used are $T=0.01$, $K=1000$, $\Lambda=1$, $U=269.35$, corresponding to $T_{sc}=1$ and $\Delta=8.39$.}
\label{fig:gapspectral}
\end{figure}

{\it Incoherent metals}-- 
In order to compute the transport properties of Model 2 described in Fig.~\ref{fig:transportgap2}, we need to numerically find the real-time Green's function $G_R(\omega)$ in the metallic phase. This can be done by solving the Dyson equation~(\ref{DysonNormal}) on the Keldysh contour. It reads as 
\begin{align}
&\Sigma_R(t) = -\frac{J^2}{128}\theta(t)\Big[(G_K(t)+G_R(t))^4(G^\ast_K(t) -G^\ast_R(t))^3-(G_K(t)-G_R(t))^4(G^\ast_K(t)+G^\ast_R(t))^3\Big] \nn
&-\frac{K^2}{8}\theta(t)\Big[(G_K(t)+G_R(t))^2(G^\ast_K(t) -G^\ast_R(t))-(G_K(t)-G_R(t))^2(G^\ast_K(t)+G^\ast_R(t))\Big], \nn
&\Sigma_K(t>0) = -\frac{J^2}{128}\Big[(G_K(t)+G_R(t))^4(G^\ast_K(t) -G^\ast_R(t))^3+(G_K(t)-G_R(t))^4(G^\ast_K(t)+G^\ast_R(t))^3\Big] \nn
&-\frac{K^2}{8}\Big[(G_K(t)+G_R(t))^2(G^\ast_K(t) -G^\ast_R(t))+(G_K(t)-G_R(t))^2(G^\ast_K(t)+G^\ast_R(t))\Big], \nn
&\Sigma_K(t<0) = -\Sigma_K^\ast(-t>0), \nn
&G_R(\omega) = \frac{2\tanh(\omega/(2T))}{2\tanh(\omega/(2T))(\omega-\mathrm{Re}[\Sigma_R(\omega)])-\Sigma_K(\omega)}, \nn
&G_K(\omega) = 2i\mathrm{Im}[G_R(\omega)]\tanh(\omega/(2T)).
\end{align}
These equations can be solved by iteration just like their imaginary time counterparts, determining $G_R(\omega)$ and $G_K(\omega)$. The real-time retarded version of the current-current correlator~(\ref{JJ2}) is
\begin{align}
&\ll \langle I^{\hat{l}} I^{\hat{l}}\rangle_R({\bf q}=0,t) \gg = \frac{NK^2}{4z}\theta(t)\Big[(G_K(t)+G_R(t))^2(G^\ast_K(t) -G^\ast_R(t))^2-(G_K(t)-G_R(t))^2(G^\ast_K(t)+G^\ast_R(t))^2\Big],
\end{align}
with the uniform DC conductivity
\beq
\sigma_{\mathrm{DC}}=\lim_{\omega\rightarrow0}\frac{d}{d\omega}(\ll \langle I^{\hat{l}} I^{\hat{l}}\rangle_R({\bf q}=0,\omega) \gg).
\eeq

\section{Gap equations for Model 2}

Since the leading instability is to a uniform paired state, after condensing the order parameter $\Delta_0 = \langle\sum_i a_{im}b_{im}\rangle/N$, we obtain the following gap equations in the large-$N$ limit:
\begin{align}
&\mathcal{G}(i\omega_n) = \frac{G(i\omega_n)}{1+U^2|\Delta_0|^2|G(i\omega_n)|^2}, ~~\Sigma(\tau-\tau^\prime) = - J^2\mathcal{G}^4(\tau-\tau^\prime)\mathcal{G}^3(\tau^\prime-\tau) -K^2\mathcal{G}^2(\tau-\tau^\prime)\mathcal{G}(\tau^\prime-\tau), \nn
&G^{-1}(i\omega_n) = i\omega_n - \Sigma(i\omega_n), ~~T\sum_{\omega_n}\frac{|G(i\omega_n)|^2\Delta_0}{1+U^2 |\Delta_0|^2 |G(i\omega_n)|^2} = \frac{\Delta_0}{U}.
\label{Dyson}
\end{align}
Sending $\Delta_0\rightarrow0$ allows us to determine $T_{sc}$.

\end{widetext}

\end{document}